\documentclass[11pt,twoside]{article}

\usepackage{asp2006}
\usepackage{epsf}
\usepackage{psfig}
\usepackage{lscape}
\usepackage{graphicx}

\begin{document}
\title{High energy X-ray emission from recurrent novae in quiescence: T CrB}   
\author{Gerardo J. M. Luna\altaffilmark{1}, J. L. Sokoloski\altaffilmark{2} \& Koji Mukai\altaffilmark{3}}   
\altaffiltext{1}{\scriptsize Instituto de Astronomia, Geofisica e Ciencias Atmosf\^ericas, Universidade de S\~ao Paulo, Rua do Mat\~ao 1226, Cid. Universitaria, S\~ao Paulo, Brazil 05508-900}
\altaffiltext{2}{\scriptsize Columbia Astrophysics Lab. 550 W120th St., 1027 Pupin Hall, Columbia University, New York, New York 10027, USA}
\altaffiltext{2}{\scriptsize CRESST and X-ray Astrophysics Laboratory NASA/GSFC, Greenbelt, MD 20771, USA and Department of Physics, University of Maryland, Baltimore county, 1000 Hilltop Circle, Baltimore, MD 21250, USA}


\begin{abstract} 
We present 
Suzaku X-ray observations of the recurrent nova T CrB in quiescence.  T CrB is the first 
recurrent nova to be detected in the hard-X-ray band ($E \sim 40.0$ keV) during quiescence. The X-ray spectrum is consistent with cooling-flow emission emanating from an optically thin region in the boundary layer of an accretion disk around the white dwarf.
The detection of strong stochastic flux variations in the light curve
supports the interpretation of the hard X-ray emission as emanating from a boundary layer. 
\end{abstract}

\section{Introduction \label{sec:intro}}

Symbiotic stars are binary systems in which a compact object accretes from red giant, the wind  of which forms a dense nebula around the binary.  The nebula is ionized by the hot compact source, which is usually a white dwarf \citep{kenyon}.  Among the known symbiotic stars \citep{belczinsky00}, four objects have shown recurrent nova (RN) eruptions: T CrB, RS Oph, V3890 Sgr and V745 Sco. These eruptions are 
triggered by a thermonuclear runaway on the white dwarf (WD) surface after accretion of a critical 
amount of H rich material from the companion. 

\citet{orio2001} suggested that during quiescence, the X-ray emission from RN could originate in the boundary layer of an accretion disc that was reconstructed around the WD after the outburst. Therefore, X-ray observations should allow one to determine a lower limit for the accretion rate.
The low X-ray luminosity measured in 350 novae and recurrent novae in quiescence \citep{orio2001} imply a low accretion rate, inconsistent with the observed recurrence times. For example, in the case of RS Oph \citep{orio1993}, the short recurrence time of $\sim$20 years requires an accretion rate of $\sim$10$^{-8}$ M$_{\odot}$/yr, inconsistent with measured X-ray fluxes. This inconsistency is known as the ``missing boundary layer problem".

T CrB is a well known recurrent nova with recorded outbursts in 1866 and 1946. 
Its quiescent-state hard X-ray emission (E$_{max}$ $\sim$ 100 keV) was first revealed during a 9 month $Swift$ survey \citep{tueller05}. Two additional $Swift$/XRT observations were analyzed by \citet{jamie}, who reported highly absorbed emission that was well fit with a cooling flow model with $kT_{max}$=23.1 keV.  The absorber consisted of components that both fully covered and partially covered the hard
X-ray source.  Here we describe Suzaku X-ray Imaging Spectrometer (XIS) and Hard X-ray Detector (HXD) observations of the recurrent nova T CrB performed on September 6, 2006. 

%

\section{Observations and data reduction \label{sec:reduc}}

On September 6th 2006, the Astro-E2/Suzaku X-ray observatory performed a 46 ks (GO) observation of T CrB (ObsId 401043010 start time  22:44:21 UT). 
We reduced the data according to standard procedures\footnote{\tiny http://www.astro.isas.jaxa.jp/suzaku/analysis/} using the software package HEASOFT 6.2\footnote{\tiny http://heasarc.gsfc.nasa.gov/docs/software/lheasoft/}. 
After determining the XIS background count rate from a nearby blank region, we extracted spectra and light curves for the XIS and HXD detectors. 
We extracted two sets of light curves for each detector -- one with a bin size of 360 s (used to search for stochastic bright variations) and the other grouping photons every 8 s. 
We used the light curve with the smaller time bins for power spectrum analysis.

\section{Analysis and Results \label{sec:spect}}

\subsection{Spectral analysis}

\begin{figure}
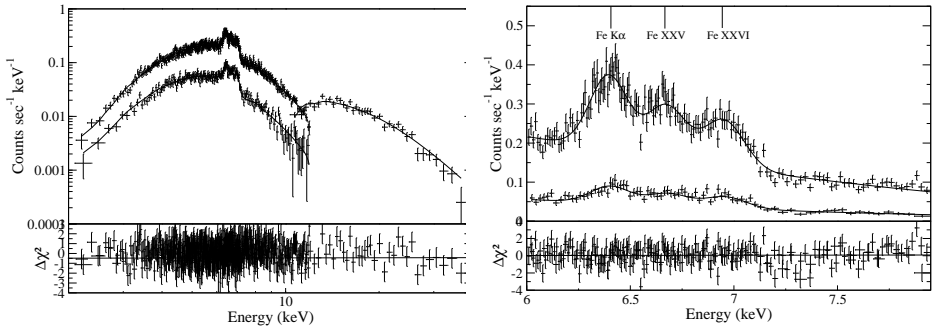

\includegraphics[angle=-90,scale=0.5]{luna_fig1.epsi}
\includegraphics[angle=-90,scale=0.5]{luna_fig2.epsi}
\caption{{\it Left panel}: Suzaku  background-subtracted spectra of T CrB between 0.3 and 40.0 keV.   
The top spectrum on the low-energy side is the sum of spectra from the front-side illuminated XIS detectors (0, 2, and 3).
The bottom spectrum on the low-energy side is from the back-side illuminated XIS detector (1).  The high-energy spectrum is from the HXD.
The best-fit absorbed cooling flow model is overplotted. 
{\it Right panel}: Fe line region ($\sim$ 6.5 keV), with best-fit {\it powerlaw} continuum plus 3 {\it gaussian} profiles overplotted. The botton panels show the residuals with respect to the models. }
\label{fig:spectrum}
\end{figure}

Visual inspection of the extracted spectrum showed prominent Fe lines around $\sim$6.5 keV.
As these lines are an indication of the presence of thermal emission from a plasma with temperatures $\sim$10$^{7-8}$ K, we first attempted to fit the
spectrum with a single-temperature plasma plus neutral absorption.  But no satisfactory fit was obtained. Even using more complex absorbers, such as an absorber that only partially covers the source, or absorption from an
 ionized plasma, did not produce acceptable fits. We did, however, find acceptable fits ($\chi^2$=1.07) for
a multi-temperature, cooling flow plasma model ($mkcflow$).  The model required a complex absorber consisting of one absorbing system that fully covered the source [n$_{H}(\rm full)=17.8_{16.6}^{19.2}\times10^{22}$ cm$^{-2}$, which includes interstellar absorption of 0.047$\times10^{22}$cm$^{-2}$] and and another that only partially covered the source [n$_{H}(\rm partial)=35.6_{32.3}^{42.2}$ $\times$10$^{22}$ cm$^{-2}$, partial covering fraction of 0.68$_{0.66}^{0.72}$]. The minimum temperature of the cooling-flow model is consistent with the smallest value allowed by the {\it mkcflow} model ($kT$=0.0808 keV), and the maximum temperature is $kT$=57.4$_{48.0}^{68.3}$ keV.  We also added a gaussian emission-line profile to the model to account for the Fe K$\alpha$ fluorescence line. Figure \ref{fig:spectrum} shows the spectra from the XIS and HXD detectors, with the cooling-flow model overplotted.  
To estimate the parameters for the Fe lines, we used a simple powerlaw fit to establish the continuum level, and then fit the Fe K$\alpha$, Fe XXV and Fe XXVI emission lines  with three Gaussian profiles (see fig. \ref{fig:spectrum}, right panel). The equivalent widths obtained were 185$_{183}^{187}$, 135$_{134}^{137}$ and 151$_{149}^{153}$ eV respectively.

\subsection{Timing analysis}

We examined the XIS 
time series binned at 360 s (Figure 2) and found significant stochastic variability above the level expected from Poisson noise. The fractional amplitude of the stochastic variations, represented as the ratio between the measured rms variation, $s$, and that expected from Poisson fluctuations alone, $s_{exp}$, is 3.75. In the case of HXD light curve, no stochastic variations were found above the Poisson level. 
Although by eye there appears to be a periodic (or quasi-periodic) variation in the
 light curve at a period of around 100 minutes, we did not detect a statistically
significant oscillation at this period in our preliminary Lomb-Scargle
analysis (taking into account the underlying broadband continuum
power) of the XIS and HXD light curves.

\begin{figure}[h!]
\begin{minipage}{0.4\textwidth}
\includegraphics[scale=0.45, angle=-90]{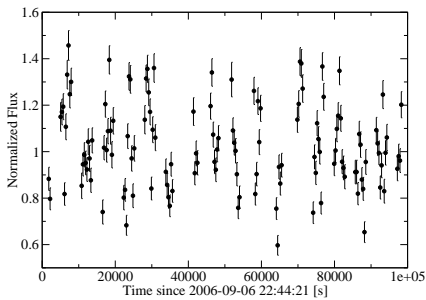}
\end{minipage}
\begin{minipage}{0.6\textwidth}
\caption{Mean normalized XIS  light curve (all four detectors added) with a 360 s bin size. The plot shows the flux as a function of time in the energy range 0.3-12.0 keV. The X-ray emission is clearly variable on a time scale of minutes.}
\end{minipage}
\end{figure}

\section{Discussion \label{sec:disc}}

Hard X-ray emission with energy up to 100 keV is observed from the 
dwarf nova SS Cyg in quiescence \citep{barlow2006}. Dwarf novae observed with XMM-$Newton$ in quiescence also display spectra with energies up to $\sim$12 keV \citep{pandel}.  For low accretion rates ($<$ 3$\times$10$^{-9}$ M$_{\odot}$/yr), \citet{narayan} showed that the most internal part of an accretion disc around a 1 M$_{\odot}$ WD, the boundary layer, radiates as an optically thin plasma at $\sim$ 10 keV. The hardness of the X-ray emission detected from T CrB strongly suggest that it comes
from the accretion region closest to the WD. From the spectral model fitted, the accretion rate is $\dot{M}$ $\sim$4.2$\times$10$^{-9}$ M$_{\odot}$ yr$^{-1}$ ($d$/1 kpc)$^{2}$.
The detection of rapid flickering, which typically emanates from regions close to the WD, supports an accretion origin for the detected emission. Moreover, the cooling-flow spectral model, as with boundary layer emission from non-magnetic cataclysmic variables \citep[CVs;][]{mukai2003} and other 
accreting high-mass white dwarfs \citep{luna2007a,smith2007}, provides a natural context for the flickering.


The EW of the neutral Fe K$\alpha$ line suggests that the X-ray source is surrounded by large amounts of neutral material \citep{inoue}. Reflection of radiation off of the WD surface can also contribute to the formation of this line. From \citet{george}, the EW of the Fe K$\alpha$ line when a 2$\pi$ source reflects bremsstrahlung photons with $\sim$10 keV is 100-200 eV. Therefore the intense Fe K$\alpha$ line is consistent either with large amounts of neutral material, as with RT Cru and CD -57 3057 \citep{luna2007a,smith2007,jamie}, or with reflection from the surface of the WD. 


\section{Conclusions \label{sec:concl}}

Using the Suzaku X-ray satellite, we observed hard-X-ray emission from the recurrent nova T CrB in quiescence. Assuming that a more detailed timing analysis confirms the lack
of periodic variations, the presence of flickering on a timescale
of minutes and the cooling-flow type spectrum suggest that the
accretion onto the WD proceeds through a disc with an optically
thin boundary layer.
T CrB is the first recurrent nova to show hard-X-ray emission from a boundary layer during quiescence. 



\acknowledgements 
G. J. M. Luna acknowledge CNPq and CAPES/PROEX support. J. L. Sokoloski
acknowledge NASA (award number NNX06AI16G) in support of the Suzaku observation of T CrB.



\end{document}